\documentclass[10pt,a4paper,twocolumn,superscriptaddress,aps,nofootinbib, showkeys, showpacs,  ]{revtex4-2}
\usepackage{amsmath,amssymb,graphicx}
\makeatletter
\usepackage[active]{srcltx}
\usepackage{graphicx,color}
\usepackage{changebar}
\usepackage{hyperref}
\hypersetup{
	colorlinks=true,
	urlcolor= blue,
	citecolor=blue,
	linkcolor= blue,
	bookmarks=true,
	bookmarksopen=false,
}
\usepackage[T1]{fontenc}
\usepackage{ae}
\usepackage[ansinew]{inputenc}
\usepackage{amsmath}
\usepackage{braket}
\usepackage{mathtools}
\usepackage{booktabs}
\usepackage{slashed}
\usepackage{empheq}
\usepackage{tikz}
\usepackage{multirow}
\usetikzlibrary{decorations.pathmorphing}
\usepackage{amsmath,amssymb,graphics,epsfig,subfigure}
\usetikzlibrary{quotes,angles}
\usepackage{etoolbox}
\usepackage{orcidlink}
 \usepackage{soul}
\usepackage[normalem]{ulem}

\usepackage{graphicx}
\usepackage{amsfonts}
\usepackage{amsmath}
\usetikzlibrary{arrows} 
\usetikzlibrary{decorations.markings}

\begin{document}

\title{Light deflection and gravitational lensing effects in acoustic black-bounce spacetime}

 \author{C. F. S. Pereira \orcidlink{0000-0001-6913-0223}}
\email{carlosfisica32@gmail.com}
\affiliation{Departamento de F\'isica e Qu\'imica, Universidade Federal do Esp\'irito Santo, Av.Fernando Ferrari, 514, Goiabeiras, Vit\'oria, ES 29060-900, Brazil}
\author{A. R. Soares \orcidlink{0000-0003-1871-2068}}
\email{adriano.soares@ifma.edu.br}

\affiliation{Grupo de Estudos e Pesquisas em Laborat\'orio de Educa\c{c}\~ao Matem\'atica, Instituto Federal de Educa\c{c}\~ao Ci\^encia e Tecnologia do Maranh\~ao,  R. Dep. Gast\~ao Vieira, 1000, CEP 65393-000 Buriticupu, MA, Brazil.}

\author{M. V. de S. Silva \orcidlink{0000-0002-8080-9277}}
	\email{marcosvinicius@fisica.ufc.br}
	\affiliation{Departamento de F\'isica, Programa de P\'os-Gradua\c c\~ao em F\'isica, Universidade Federal do Cear\'a, Campus Pici, 60440-900, Fortaleza, Cear\'a, Brazil}	
\author{R. L. L. Vit\'oria \orcidlink{0000-0001-8802-3634}}
\email{ricardovitoria@professor.uema.br/ricardo-luis91@hotmail.com}
\affiliation{Departamento de Ci\^encia Exatas e Naturais, Universidade Estadual do Maranh\~ao, Rua Dias Carneiro, Contorno da Avenida Jo\~ao Alberto de Sousa, s/n, Ramal, Bacabal, MA 65.700-000, Brazil}
\affiliation{Centro de Ci\^encias Humanas, Naturais, Sa\'ude e Tecnologia, Universidade Federal do Maranh\~ao, Estrada Pinheiro/Pacas, Km 10, s/n, Enseada, Pinheiro, MA 65.200-000, Brazil}
\author{H. Belich  \orcidlink{0000-0002-8795-1735}}
\email{humberto.belich@ufes.br}
\affiliation{Departamento de F\'isica e Qu\'imica, Universidade Federal do Esp\'irito Santo, Av.Fernando Ferrari, 514, Goiabeiras, Vit\'oria, ES 29060-900, Brazil.}

	
\begin{abstract}

    In the present work, we analyze the gravitational deflection for a light beam in the weak and strong field regimes for the gravitational analog geometry of an acoustic black hole (ABH) and acoustic black-bounce (ABB). Motivationally, the first spacetime arises as an exact solution of the field equations for gravitational black holes (BHs) in Einstein-scalar-Gauss-Bonnet theory (EsGB) P. Ca\~nate et al. $ \left( Class.~ Quant.~ Grav. ~\textbf{38}, 125002~ (2021)\right)$. In contrast, the second model arises from the combination of phantom scalar field and nonlinear electrodynamics in general relativity (GR) P. Ca\~nate  $ \left( Phys. ~Rev. D ~\textbf{106}, no.2, 024031 ~(2022)\right)$. We construct analytical expressions for the angular deflection of light in both limits and, from them, analyze the construction of the observables, which allow us to relate theoretical models to observational data. We compare these observables and show how much they differ from those obtained in the Schwarzschild solution.
	
\end{abstract}


\maketitle

\section{Introduction}\label{sec1}

Singular BH solutions arise naturally in the context of GR theory through the so-called gravitational collapse as the final phase of a supermassive star, with the first and simplest proposal for this object being presented by Karl Schwarzschild representing a vacuum solution to Einstein's theory \cite{INTROBORD1,INTROBORD2,INTROBORD3,INTROBORD4}. Later, other solutions were proposed, such as the case of the electrically charged spacetime of Reissner-Nordstr\"om representing the electrovacuum, models with cosmological constant and rotation scenarios such as the case of Kerr and Kerr-Newman \cite{INTROBORD1,INTROBORD2}. To remove the existence of the singularity present inside a BH, the first model of a regular BH was proposed around 1968 by James Bardeen \cite{INTROBORD5}. Such spacetime did not satisfy Einstein's equations in vacuum \cite{INTROBORD5,INTROBORD6}, and therefore, it was only around the year 2000 that its respective matter content was obtained by Beato and Garcia \cite{INTROBORD7}. The matter content was formed by a class of nonlinear electrodynamics, in which the regularization parameter is identified as being the charge of the magnetic monopole. The electrically charged version of Bardeen's spacetime was developed numerically by Rodrigues and Silva in \cite{INTROBORD6}. In this way, the scenario of regular BH solutions has been explored in the most varied contexts \cite{INTROB21,INTROB22,INTROB23,INTROB24,INTROB25,INTROB26,INTROB27,INTROB28,INTROB29,INTROB210,INTROB211,INTROB212,INTROB213,INTROB214}.

Recently, further enriching the scenario of regular solutions, Simpson and Visser proposed a new class of solutions called black bounce (BB) \cite{7}. In a very simplified way, the model structure differs from a regular BH due to the presence of an area function that hides the presence of the singularity or otherwise represents the non-zero throat of the wormhole (WH). In the present model, there are some possibilities of formation of structures according to the way in which the mass parameters $m$ and the throat $a$ are related: for $a>2m$, there is the formation of a bidirectional WH, for $a=2m$ we have a unidirectional WH with the throat located at the origin $r=0$, and for $a<2m$, we have the formation of a WH with the presence of two symmetrical horizons. Similarly to Bardeen's model, this BB spacetime initially does not satisfy Einstein's equations; it was only later that Bronnikov and Walia obtained the matter content associated with such spacetime, being composed of a class of nonlinear electrodynamics and phantom scalar field \cite{INTRO22}. Since then, these BB models have been widely explored in a wide variety of scenarios \cite{INTRO24,INTROB31,INTROB32,INTROB33,INTROB34,INTROB35,INTROB36,INTROB37,INTROB38,INTROB39,INTROB310,INTROB311,INTROB312,INTROB313,INTROB314,INTROB315,INTROB316,INTROB317,INTROB318,INTROB319,INTROB320,INTROB321,INTROB322}.

We call gravitational lensing the phenomenon where light from a distant source is deflected by the gravitational field of a massive object, such as stars, BH, and galaxies \cite{INTRO45,INTRO46}. In the context of a BH, we can distinguish two limits. When light passes at a considerable distance from the BH, the deflection is small; we call this the weak field limit. This means that the light deviation is subtle, and the source images do not undergo extreme distortions. On the other hand, as light passes closer to the BH, approaching the photon sphere (a region where photons can orbit the BH), the deflection becomes much more intense; we call this the strong field limit. In this regime, light can make multiple orbits around the BH before being observed, and this deflection can even diverge, creating dramatic distortions. The images formed in this limit are called relativistic images. Some pioneering work on the subject was proposed in the 1960s \cite{INTRO47,INTRO48}, which showed that relativistic images were very weak and that there were no efficient mathematical methods for calculating deflection in the strong field regime. It was not until the 2000s that Virbhadra and Ellis \cite{8} formalized a method for constructing minimally consistent gravitational lensing equations in the strong field regime. Bozza \cite{5} then developed a technique for obtaining the equations in the strong field regime that was later improved by Tsukamoto \cite{6}.

Since then, the phenomenology of gravitational lenses has been explored in a wide variety of scenarios due to its possibility of measurement, contexts such as BHs \cite{KS1,INTROB51,INTROB52,INTROB53,INTROB54,INTROB55,INTROB56,KS2,INTROB57,INTROB58,INTROB59,INTROB510,INTROB511,INTROB512,INTROB513,INTROB514,INTROB515}, WHs \cite{INTROB61,INTROB62,INTROB63,INTROB64,INTROB65,INTROB66,INTROB67,INTROB68,INTROB69,INTROB610,INTROB611}, topological defects \cite{INTROB71,INTROB72,INTROB74,INTROB75,INTROB76}, modified gravity \cite{INTROB81,INTROB82,INTROB83,INTROB84,INTROB85,INTROB86,INTROB87,INTROB88,INTROB89,INTROB75}, regular BHs, and BBs \cite{INTROB91,INTROB92,INTROB93,INTROB94,INTROB95,INTROB96,INTROB97,INTROB98,INTROB99,INTROB910,INTROB911}.

A complementary way to study the characteristics of a BH is through its laboratory analogs, the so-called ABH. This line of research began with the work of Unruh, who showed that, under certain conditions, acoustic perturbations in a fluid propagate similarly to a scalar field in curved spacetime \cite{2}. Depending on the properties of the fluid, an event horizon and an ergoregion may be present \cite{1}. The study of these analog models has been extensively developed from both theoretical and experimental perspectives \cite{INTROB101,INTROB102,INTROB103,INTROB104,INTROB105,INTROB106,INTROB107,INTROB108,INTROB109,INTROB1011,INTROB1012,INTROB1013,INTROB1014,INTROB1015}.

Analog models are represented by effective metrics, which are the metrics that perturbations experience as they propagate through fluids, rather than the actual metrics that describe a curved spacetime. However, it is possible to obtain these effective metrics as solutions to the field equations of gravitational theories. In \cite{3}, the authors show that it is possible to obtain a line element with the same form as the canonical acoustic black hole metric by considering the EsGB theory. This solution is particularly compelling because it is the first fully analytic four-dimensional BH metric in EsGB theory that exactly reproduces an acoustic geometry, admitting closed-form expressions for the horizon radius and the location of the photon sphere; usually solutions of this type of theory need to be obtained numerically. Moreover, for some special cases, the scalar field remains finite at the horizon, and the energy conditions are satisfied in some regions of the spacetime, providing a self-consistent framework in which to compute, without further approximation, the precise scalar-Gauss-Bonnet-induced shifts in shadow radius, quasinormal mode spectrum, and other BH observables. Although the previous solution provides an exact BH geometry in EsGB theory that mimics the ABH, it still suffers from a curvature singularity. This naturally motivates the search for a regularized version of this gravitational analog, in the spirit of the BB models. This allows us to explore how regularity conditions, modified horizons, and possible throat structures influence observables such as the shadow radius and photon orbits. In \cite{INTRO24}, the authors showed that the canonical ABH metric regularized by the Simpson-Visser method can be obtained as a solution of the Einstein equations by considering a phantom scalar field together with nonlinear electrodynamics. In this way, the gravitational analogs of ABH are constructed.

In the present work, we apply the methodology for calculating the light deflection due to the gravitational field for two models, one of which is a BH and the other a magnetically charged BB. This allows us to theoretically construct quantities that can be measured in an observational way, such as the angular separation between the relativistic images and the magnification of these images. With this information, it is possible to corroborate or rule out gravitational theories based on observation. Given that alternative theories of gravity tend to behave like GR in the weak field limit, the strong field limit proves more suitable for investigating the plausibility of models that go beyond GR. In this sense, angular deflection is an excellent parameter for investigation. We apply the gravitational lensing formalism in the weak and strong field regimes for the ABB model and then construct the expressions referring to the observables. In a very natural way we can obtain the respective equivalent results for the ABH when we consider the bounce radius going to zero $a\to{0}$ showing the efficiency of the methodology.

The paper is organized as follows: In section \ref{sec2}, the general relations for a spherically symmetric and static spacetime as well as its conserved quantities are constructed. In sections \ref{sec3} and \ref{sec4}, we perform the calculation of the light deflection in the weak and strong field regimes, respectively, for a ABH and ABB. In section \ref{sec5}, we construct the gravitational lensing equations for the ABB only in both limits as well as the equations for observables. Finally, in section \ref{sec6}, we make the final discussions and conclusions of the work, presenting our perspectives.

\section{General relations}\label{sec2}

In this section, we derive general relations in order to simplify our discussions in the following sections. Since we investigate lensing in a static and spherically symmetric spacetime, let us introduce the generic line element in spherical coordinates $(t,r,\theta, \phi)$,
\begin{equation}\label{1}
ds^2= -f(r)dt^2 + \frac{dr^2}{f(r)} +\Sigma^2(r)\left(d\theta^2+\sin^2\theta{d\phi^2}\right),
\end{equation} where $f\left(r\right)$ and $\Sigma\left(r\right)$ are functions that depend only on the radial coordinates. Thus, we define a smooth curve in this generic spacetime Eq. (\ref{1}) that has length $S$ and is given by
\begin{equation}\label{2}
S= \int \sqrt{\left(g_{\mu\nu}\frac{dx^\mu}{d\lambda}\frac{dx^\nu}{d\lambda}\right)}d\lambda,
\end{equation} where $\lambda$ is an affine parameter that can represent the observer's own time. Taking $S$ as the affine parameter itself, we can show that the curve that minimizes Eq. (\ref{2}) is the same one that minimizes
\begin{equation}\label{3}
\int \left(g_{\mu\nu}\frac{dx^\mu}{d\lambda}\frac{dx^\nu}{d\lambda}\right)d\lambda=\int\mathcal{L}{d\lambda}.
\end{equation}

Therefore, considering the analysis in the equatorial plane, $\theta=\frac{\pi}{2}$, the Lagrangian becomes
\begin{equation}\label{4}
\mathcal{L}=-f(r)\left(\frac{dt}{d\lambda}\right)^2 + \frac{1}{f(r)}\left(\frac{dr}{d\lambda}\right)^2 + \Sigma^2(r)\left(\frac{d\phi}{d\lambda}\right)^2.
\end{equation}

Applying the Euler-Lagrange equations to the above expression, we define the quantities conserved at time $t$ and at $\phi$. Therefore, we have
\begin{equation}\label{5}
L= \Sigma^2(r)\frac{d\phi}{d\lambda},  \qquad \qquad E=f(r)\frac{dt}{d\lambda}.
\end{equation}

Thus, substituting the conserved quantities Eq. (\ref{5}) into Eq. (\ref{4}) and considering only null geodesics $\mathcal{L}=0$, leads (\ref{4}) to
\begin{equation}\label{6}
\left(\frac{dr}{d\lambda}\right)^2= E^2- \frac{L^2{f(r)}}{\Sigma^2(r)}.
\end{equation}

The above expression can be compared with the dynamics of a classical particle of unit mass with energy $\mathcal{E}$ and subject to an effective potential $V_{\text{eff}}$, given by
\begin{equation}\label{en-v}
	\mathcal{E}=E^2\qquad\text{and}\qquad V_{\text{eff}}=\frac{L^2f(r)}{\Sigma^2} \ .
\end{equation}
There is radial motion only if $\mathcal{E}>V_{\text{eff}}$. Here, we consider the scenario in which the photon starts its trajectory in an asymptotically flat region and approaches a distance $r_0$ from the center of the BH, reaching the so-called turning point that lies beyond the event horizon. Naturally, when the photon reaches this boundary region, due to the presence of the gravitational field, it tends to return to another asymptotically flat region. At the turning point, $V_{\text{eff}}(r_0)=E^2$, which from Eq. (\ref{en-v}) implies
\begin{equation}\label{key}
	\frac{1}{\beta^2}=\frac{f(r_0)}{\Sigma^2 (r_0)} \ ,
\end{equation}
where $\beta=\frac{L}{E}$ is the  impact parameter of the light ray. At the photon sphere radius, $r_m$, we have
\begin{equation}\label{esferadefotons}
	\frac{dV_{\text{eff}}}{dr}\Bigg|_{r_m}=0 \ .
\end{equation}
The critical impact parameter, $\beta_c$, is defined as $\beta_c=\beta(r_m)$. Light rays with $\beta<\beta_c$ are absorbed, those with $\beta=\beta_c$ end up in the photon sphere, whereas those with $\beta>\beta_c$ are scattered. Concerning the scattering, the deflection of light becomes divergent as $r_0\to r_m$, and this limit is referred to as the strong field limit. In the following sections, we will compute the approximation for the deflection angle in this limit for two gravitational models that have recently drawn the attention of the scientific community.

\section{Acoustic black hole}\label{sec3}

We apply the methodology regarding the deflection of light when subjected to a gravitational field to a spherically symmetric and static solution for ABH originally derived in Ref. \cite{1}. These models are not acoustic analogs of gravitational systems, where these models are formed when the fluid velocity is greater than the speed of sound, forming the so-called sonic horizons that are analogous to geometric event horizons \cite{1,2}. Recently, the gravitational scenario has been interpreted as a solution of BHs in EsGB theories \cite{3} and then extended to BB solutions \cite{INTRO24}.

For this section, we consider the model for an ABH where the metric functions Eq. (\ref{1}) are defined as
\begin{equation}\label{func}
	\Sigma^2(r)=r^2\qquad\text{and}\qquad f(r)= \left(1-\frac{q^2}{r^4}\right) \ .
\end{equation}
This geometry represents an ABH that has a singularity at the origin $r=0$ and the local acoustic horizon at $r_h=\sqrt{|q|}$. From Eq. (\ref{esferadefotons}), the radius of the photon sphere is given by $r_m=3^{1/4}\sqrt{|q|}$ where we consider for the study of lensing only the radius corresponding to $r>0$. This parameter $q$ is associated with the fluid density in the case of ABH and can be interpreted as being the magnetic charge for the BH and WH solutions in the EsGB theory using a self interacting scalar field.

\subsection{Deflection of light in the weak field limit}\label{sec31}


Considering the deflection of light in this specific scenario, whose metric functions are given by Eqs. (\ref{func}) and (\ref{key}), yields the following equation for the impact parameter of the light ray:
\begin{eqnarray}\label{7}
\frac{1}{\beta^2}=\frac{1}{{r^2_0}}\left(1-\frac{q^2}{{r^4_0}}\right).
\end{eqnarray} 
Using Eq. (\ref{func}), the radial equation, Eq. (\ref{6}), becomes
\begin{equation}\label{dr}
	\bigg(\frac{dr}{d\lambda}\bigg)^2=E^2-\frac{L^2}{r^2}\bigg(1-\frac{q^2}{r^4}\bigg) \ .
\end{equation}
Substituting Eq. (\ref{5}) in Eq. (\ref{dr}), we find
\begin{equation}
	\Big(\frac{dr}{d\phi}\Big)^2=\frac{r^4}{\beta^2}-r^2\bigg(1-\frac{q^2}{r^4}\bigg) \ .
\end{equation}

Considering that the distance before and after the turning point are symmetrical and therefore the contributions to the angular deviation are the same, we then have that 
\begin{equation}\label{8}
	\Delta\phi=\pm{2}\int_{r_0}^{\infty}\Big[\frac{r^4}{\beta^2}-r^2\bigg(1-\frac{q^2}{r^4}\bigg)\Big]^{-1/2} \ dr.
	\end{equation} Using the symmetry of the solution, we adopt only the positive sign of the expression above.

Let us introduce the change of coordinates in Eq. (\ref{8}), where $u=\frac{1}{r}$. Note that the limits of integration become $u\to{0}$ when $r\to\infty$ and $u\to{u_0}=\frac{1}{r_0}$ when $r\to{r_0}$. Thus, the equation for the angular deviation in the new coordinates is expressed by
\begin{eqnarray}\label{9}
\Delta\phi=  2\int^{u_0}_{0}du\left[\frac{1}{\beta^2(u_0)}-u^2\left(1-q^2u^4\right)\right]^{-1/2}.
\end{eqnarray}

\begin{figure}[b]
   	\centering
   	\includegraphics[width=\columnwidth]{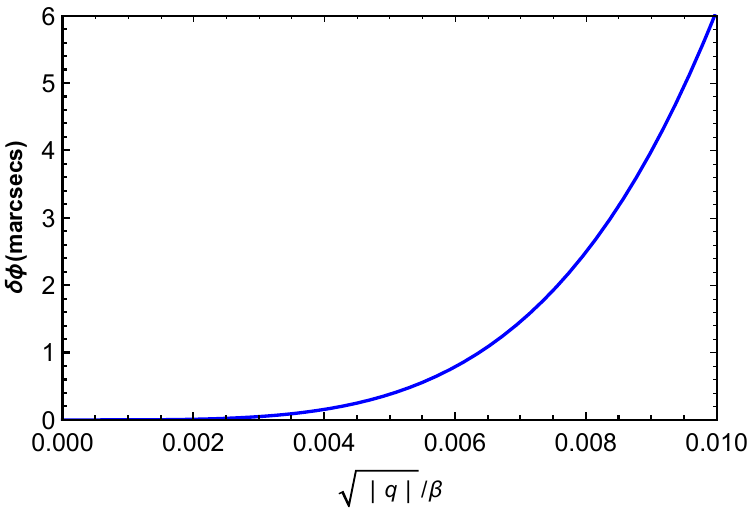}
   	\caption{Angular deflection of light in the weak field regime in milliseconds of arc.} 
   	\label{wf}
   \end{figure}

To perform the angular deviation expansion in this weak field regime, we have to return to the impact parameter expression, Eq. (\ref{7}) and then use the same coordinate change as above $\frac{1}{\beta^2}=u^2_0\left(1-q^2u^4_0\right)$. Thus, we used this new expression for the impact parameter and assumed that the photon passes very far from the BH. Then, we can consider the approximation that the magnetic charge $q\ll{1}$ and then expand the orbit expression, Eq. (\ref{9}), to fourth order in the magnetic charge $q$, and we have that the deviation of the light is given by $\delta\phi=\Delta\phi-\pi$,
\begin{eqnarray}\label{10}
\delta\phi \simeq \frac{15\pi{q^2}}{16\beta^4} + \frac{1545\pi{q^4}}{1024\beta^8}.
\end{eqnarray}
To estimate the order of magnitude of the angular deflection in the weak field limit, we plot $\delta\phi$ as a function of the ratio $\sqrt{|q|}/\beta$ in Fig.~(\ref{wf}).
As we can observe, the deflection decreases from values on the order of milliseconds of arc to zero as the ratio $\sqrt{|q|}/\beta \to 0$, which corresponds to asymptotically flat regions of this spacetime.

\subsection{Deflection of light in the strong field limit}\label{sec32}

To develop the calculation of gravitational deflection of light in the strong field regime, we use as a starting point the methodology developed by Bozza \cite{5} and improved by Tsukamoto \cite{6}. To do so, we begin by performing the following change of variables $z=1-\frac{r_0}{r}$ in the orbit equation, Eq. (\ref{8}), and then we can rewrite it as
\begin{eqnarray}\label{11}
\Delta\phi(r_0)=\pm \int^1_0 \frac{2r_0{dz}}{\sqrt{G(z,r_0)}},
\end{eqnarray} where
\begin{eqnarray}\label{12}
G(z,r_0)= \frac{r^4_0}{\beta^2} -r^2_0\left(1-z\right)^2 + \frac{q^2\left(1-z\right)^6}{r^2_0}.
\end{eqnarray}

Performing the series expansion of the function $G(z,r_0)$ at the point where $z=0$, which is equivalent to the limit of $r\to {r_0}$, we have that
\begin{eqnarray}\label{13}
G(z,r_0) \simeq \Lambda_1(r_0)z + \Lambda_2(r_0)z^2,
\end{eqnarray} where the expansion parameters $\Lambda_1$ and $\Lambda_2$ are defined as
\begin{equation}\label{14}
\Lambda_1(r_0)=\frac{2}{r^2_0}\left(r^4_0 - 3q^2\right), \qquad \Lambda_1(r_0\to{r_m})=0, 
\end{equation}
\begin{eqnarray}\label{15}
\Lambda_2(r_0)=\frac{1}{r^2_0}\left(15q^2-r^4_0\right), \\\label{15a}
\Lambda_2(r_0\to{r_m})= 4\sqrt{3}|q|.
\end{eqnarray}

The expansion coefficients Eqs. (\ref{14}) and (\ref{15a}) in the strong field regime, $r\to {r_m}$, cause the integral Eq. (\ref{11}) to have a logarithmic divergence, and therefore, it is necessary to separate this integral into two parts: a regular part $\Delta\phi_R(r_0)$ and a divergent part $\Delta\phi_D(r_0)$. Therefore, we have
\begin{eqnarray}\label{16}
\Delta\phi(r_0) = \Delta\phi_R(r_0) + \Delta\phi_D(r_0).
\end{eqnarray}

Thus, the divergent part is defined by
\begin{eqnarray}\label{17}
\Delta\phi_D(r_0)&=& \int^1_0\frac{2r_0}{\sqrt{\Lambda_1(r_0)z+\Lambda_2(r_0)z^2}}= \nonumber \\
&-&\frac{4r_0}{\sqrt{\Lambda_2(r_0)}}\log\left(\sqrt{\Lambda_1(r_0)}\right) \nonumber \\
&+& \frac{4r_0}{\sqrt{\Lambda_2(r_0)}}\log\left(\sqrt{4\Lambda_2(r_0)}\right). \nonumber \\
\end{eqnarray}

To have some control over the divergent part, we need to expand the coefficient $\Lambda_1(r_0)$ and the impact parameter $\beta(r_0)$ in the radius of the photon sphere $r_m$. Thus, Eqs. (7) and (14) become
\begin{equation}\label{18}
    \beta\left(r_0\right) \simeq  3^{3/4}\sqrt{\frac{|q|}{2}} +\frac{\left(r^4_0-3q^2\right)^2}{\left(2^{9/2}{3}^{1/4}{|q|}^{7/2}\right)},
\end{equation}
\begin{equation}\label{19}
\Lambda_1(r_0) \simeq \frac{2}{\sqrt{3}|q|}\left[\left(2^{9/2}3^{1/4}|q|^{7/2}\right)\left(\beta - 3^{3/4}\sqrt{\frac{|q|}{2}}\right)\right]^{1/2}.
\end{equation}

Thus, substituting Eqs. (\ref{15a}), (\ref{18}), and (\ref{19}) in the expression referring to the divergent part, Eq. (\ref{17}), we see that
\begin{eqnarray}\label{20}
\Delta\phi_D= -\frac{1}{2}\log\left[\frac{\beta}{3^{3/4}\sqrt{\frac{|q|}{2}}}-1\right] + \frac{\log\left(12\right)}{2}.
\end{eqnarray}

The equation above represents the analytical expression referring to the divergent part of the deviation of light in the spacetime of an ABH. Regarding the contribution related to the regular part of the gravitational deviation, we have to consider the limit in which the impact parameter, Eq. (\ref{7}), assumes its critical value $r_0\to{r_m}$ and then use Eqs. (\ref{12}) and (\ref{15a}) in this limit. Thus, the general expression of the regular part becomes
\begin{equation}\label{21}
\Delta\phi_R= \int^{1}_0 \frac{2r_{m}}{\sqrt{G(z,r_{m})}}dz 
-\int^{1}_{0}\frac{2r_{m}}{\sqrt{\Lambda_2(r_{m})}}\frac{dz}{z}=\log(3).
\end{equation}

Therefore, the total deviation of the light suffered due to the presence of the gravitational field in this ABH is defined as $\delta\phi=\Delta\phi_D + \Delta\phi_R -\pi$, and its expression can be obtained analytically, as shown above. The interpretation of the single parameter of the model $q$ can follow the line of being associated with the density of the fluid in the acoustic case or as the magnetic charge in EsGB theory.

\section{Acoustic black bounce}\label{sec4}

In this section, we study the deflection of light in the weak and strong field regime using as a background an ABB model that was derived in Ref. \cite{INTRO24} initially considering an ABH and then applying the Simpson and Visser regularization method \cite{7}. In the gravitational context, this solution arises from Einstein equations when a phantom scalar field is considered together with a nonlinear electrodynamics. Thus, for better comparisons with Ref. \cite{INTRO24}, the radial coordinate defined in the first section of this work is now to be described by the variable $\rho$, and the metric functions are expressed by
\begin{equation}\label{22}
    \Sigma^2(\rho)= \rho^2+a^2, \qquad f(\rho)=1-\frac{q^2}{\left(\rho^2+a^2\right)^2} \ ,
\end{equation} where the model parameters are now represented by the magnetic charge $q$ and the WH throat radius $a$. The position of the event horizon and the radius of the photon sphere are defined by $\rho_H= \sqrt{|q|-a^2}$ and $\rho_m=\left(\sqrt{3}|q|-a^2\right)^{1/2}$. 

In the present work, we consider only the configurations in which $|q|>a^2$  represents the spacetime of an ABB. Other possible structures can be found in Ref. \cite{INTRO24}.

\subsection{Expansion for light deflection in the weak field limit}\label{sec41}

Just as in the first model, we consider in the weak field regime a photon that starts its trajectory in an asymptotically flat region and approaches the center of the ABB at a distance $\rho_0$ defined as the turning point. Based on Eq.~(\ref{key}), the expression for the impact parameter in this spacetime, given in Eq.~(\ref{22}), takes the form
\begin{eqnarray}\label{23}
    \frac{1}{\beta^2}= \frac{1}{\left(\rho^2_0 + a^2\right)}\left(1-\frac{q^2}{\left(\rho^2_0 + a^2\right)^2}\right).
\end{eqnarray}

From Eq.~(\ref{6}), we obtain the following radial equation:
\begin{equation}\label{dr2}
	\bigg(\frac{d\rho}{d\lambda}\bigg)^2 = E^2 - \frac{L^2}{(\rho^2 + a^2)}\Big[1 - \frac{q^2}{(\rho^2 + a^2)^2}\Big] \ .
\end{equation}
From Eq.~(\ref{5}), Eq.~(\ref{dr2}) can be rewritten as
\begin{equation}\label{dr22}
	\bigg(\frac{d\rho}{d\phi}\bigg)^2 = \frac{(\rho^2 + a^2)^2}{\beta^2} - (\rho^2 + a^2)\Big[1 - \frac{q^2}{(\rho^2 + a^2)^2}\Big] \ .
\end{equation}

Considering the symmetry of the system with respect to the position of the turning point, the expression for the angular deviation is given by
\begin{eqnarray}\label{24}
\Delta\phi &=& \pm{2}\int^{\infty}_{\rho_0} \left[\frac{(\rho^2+a^2)^2}{\beta^2}\right. \nonumber\\
&&\left. -(\rho^2+a^2)\Big(1-\frac{q^2}{(\rho^2+a^2)^2}\Big)\right]^{-1/2} \ d\rho \ .
\end{eqnarray}
Using the symmetry of the solution, we adopt only the positive sign of the expression above.

Let us introduce the following change of variables in the expression above $u=1/\sqrt{\rho^2+a^2}$. Naturally, we should recover the results of the first model in the limit where the WH throat radius tends to zero $a\to{0}$. Note that the integration limits in the new take the form of $u\to{0}$ when $\rho\to{\infty}$ and $u\to{u_0}=1/\sqrt{\rho^2_0 +a^2}$ when $\rho\to{\rho_0}$.

Thus, the equation for the angular deviation in the new
coordinates is expressed by

\begin{eqnarray}\label{25}
\Delta\phi&=& 2\int^{u_0}_{0}\left[(1-u^2a^2)\right.\nonumber\\
&&\left.\times\left(\frac{1}{\beta^2(\rho_0)}-u^2+q^2u^6\right)\right]^{-1/2} \ du \ .
\end{eqnarray}

To perform the light deflection expansion in the weak field regime, it is necessary to write the impact parameter, Eq. (\ref{23}), in the new variables $\frac{1}{\beta^2}=u^2_0(1-q^2u^4_0)$. Therefore, we consider the expansion of the impact parameter in the gravitational deflection of the light equation and assume that the photon is very far from the WH. We can then perform the following approximations on the model parameters $q\ll{1}$ and $a\ll{1}$ and then write the angular deflection $\delta\phi=\Delta\phi-\pi$:
\begin{eqnarray}\label{26}
\delta\phi\simeq \frac{\pi{a^2}}{4\beta^2} + \frac{15\pi{q^2}}{16\beta^4} +\frac{9\pi{a^4}}{64\beta^4}+ \frac{19\pi{a^2q^2}}{64\beta^6} + \frac{1545\pi{q^4}}{1024\beta^8}. \nonumber \\
\end{eqnarray}
Note that in the limit where $a\to{0}$ we recover the results regarding the deviation of light in an ABH, Eq. (\ref{10}). 

\begin{figure}[ht]
   	\centering
   	\includegraphics[width=\columnwidth]{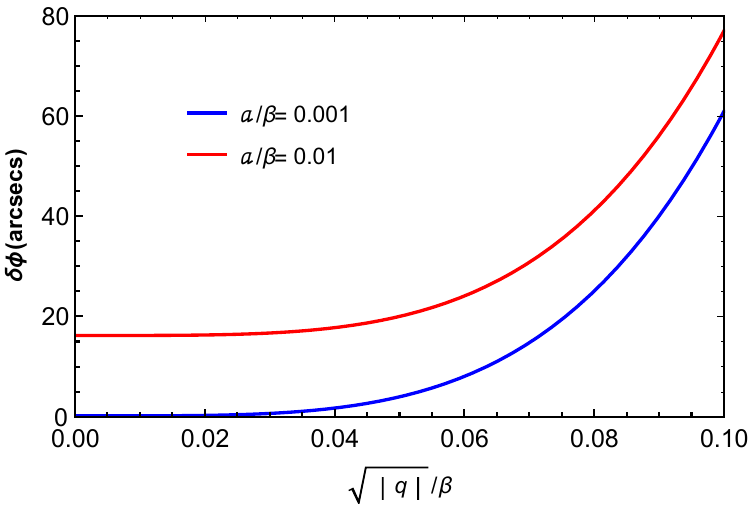}
   	\caption{Angular deflection in seconds of arc for some values of $a/\beta$, as a function of $\sqrt{|q|}/\beta$.} 
   	\label{sf}
   \end{figure}

In order to provide a visual representation of the angular deflection in the weak field limit, we constructed the plot of $\delta\phi$ for small values of the ratios $\sqrt{|q|}/\beta$ and $a/\beta$, since in this limit the light passes far from the source, see Fig.~(\ref{sf}). We can observe that the angular deflection decreases as $\sqrt{|q|}/\beta$ and $a/\beta$ become increasingly smaller, which is what occurs in the asymptotic regime.

On the other hand, when we consider the limit in which the magnetic charge of the ABB tends to zero $q\to{0}$, we recover the result regarding the angular deviation of the Ellis-Bronnikov WH \cite{r4a,r4b}, which is written as
\begin{equation}
    \delta\phi\simeq \frac{\pi{a^2}}{4\beta^2} +\frac{9\pi{a^4}}{64\beta^4}.
\end{equation}
\subsection{Deflection of light in the strong field limit}\label{sec42}

Here, as in the first model, we carry out a study related to the strong field regime using the same methodology applied previously. To do so, we consider the following change of variables, $z=1-\rho_0/\rho$, in Eq. (\ref{24}), and then, we can rewrite it as
\begin{eqnarray}\label{27}
\Delta\phi(\rho_0)=\pm \int^{1}_0 \frac{2\rho_0{dz}}{\sqrt{G(z,\rho_0)}},
\end{eqnarray} where
\begin{eqnarray}\label{28}
G(z,\rho_0)= \left[\rho^4_0+a^4(1-z)^4+2a^2\rho^2_0(1-z)^2\right] \nonumber \\
\times \left[\frac{1}{\beta^2} - \frac{(1-z)^2}{(\rho^2_0+a^2(1-z)^2)} +\frac{q^2(1-z)^6}{(\rho^2_0+a^2(1-z)^2)^3}\right].\nonumber\\
\end{eqnarray}

Performing the power series expansion for the function $G(z,\rho_0)$ around the point $z=0$, which is equivalent to considering the limit at which $\rho_0\to\rho_m$, then we have that
\begin{eqnarray}\label{29}
    G(z,\rho_0) \simeq \Lambda_1(\rho_0)z + \Lambda_2(\rho_0)z^2,  
\end{eqnarray} where the expansion parameters $\Lambda_1$ and $\Lambda_2$ are defined as
\begin{eqnarray}\label{30}
\Lambda_1(\rho_0)&=& 2\rho^2_0 \left[1-\frac{3q^2}{(\rho^2_0 +a^2)^2}\right], \\\label{31}
\Lambda_2(\rho_0)&=& - \frac{\rho^2_0\left[\rho^4_0 +5a^4+6a^2\rho^2_0-15q^2\right]}{(\rho^2_0 +a ^2)^2}.
\end{eqnarray}

In the limit where the expansion parameters tend to $\rho_0\to\rho_m$, they become
\begin{eqnarray}\label{32}
    \Lambda_1(\rho_0\to\rho_m)&=&0, \\\label{33}
    \Lambda_2(\rho_0\to\rho_m)&=& 4\sqrt{3}|q| + \frac{4a^4}{\sqrt{3}|q|} -8a^2.
\end{eqnarray}

Just as analyzed in the first model when we use the expansion coefficients in the strong field limit, $\rho_0\to\rho_m$, the expression for the angular deviation, Eq. (\ref{27}), has a logarithmic divergence, and therefore, we are led to separate it into two parts, one with the divergence and the other with the regular $\Delta\phi(\rho_0)=\Delta\phi_D(\rho_0)+\Delta\phi_R(\rho_0)$.  

Therefore, the divergent part in general form is given by Eq. (\ref{17}). Before we explicitly write the expression for the divergent part, we must address how the divergence occurs. Therefore, we need to write the expansion parameter $\Lambda_1(\rho_0)$ in terms of the impact parameter $\beta(\rho_0)$ in the limit where $\rho_0\to\rho_m$. Thus, combining the expressions Eq. (\ref{23}) and (\ref{30}), we have
\begin{eqnarray}\label{34}
    \beta(\rho_0) \simeq  3^{3/4}\sqrt{\frac{|q|}{2}} + \frac{\left[\left((\rho^2_0 +a^2)^2-3q^2\right)^2\right]}{(2^{9/2}3^{1/4}|q|^{7/2})}, \\\label{35}
    \Lambda_1(\rho_0) \simeq \frac{2(\sqrt{3}|q|-a^2)}{3q^2} \nonumber \\
    \times\left[(2^{9/2}3^{1/4}|q|^{7/2})^{1/2}(\beta - 3^{3/4}\sqrt{\frac{|q|}{2}})^{1/2}\right].
\end{eqnarray}

Therefore, combining Eqs. (\ref{33}), (\ref{34}), and (\ref{35}) in the analog of the divergent contribution to this model (\ref{17}), we have
\begin{eqnarray}\label{36}
\Delta\phi_D &=& -\frac{1}{2}\left(\frac{\sqrt{3}|q|}{\sqrt{3}|q|-a^2}\right)^{1/2} \log\left[\frac{\beta}{3^{3/4}\sqrt{\frac{|q|}{2}}}-1\right] \nonumber \\
&+&\frac{1}{2}\left(\frac{\sqrt{3}|q|}{\sqrt{3}|q|-a^2}\right)^{1/2}\log\left[\frac{4(\sqrt{3}|q|-a^2)^2}{q^2}\right]. \nonumber \\
\end{eqnarray}


The expression above represents the contribution related to the divergent part of the light deflection in spacetime of an ABB in the strong field regime. Note that in the limit where the WH throat parameter ceases to act $a\to{0}$, we recover the same expression of the first model, Eq. (\ref{20}).

The expression referring to the regular part of the integration is given by
\begin{eqnarray}\label{37}
  \Delta\phi_R= \int^1_0 \frac{2\rho_m{dz}}{\sqrt{G(z,\rho_m)}}  -\int^1_0 \frac{2\rho_m{dz}}{z\sqrt{\Lambda_2(\rho_m)}}.
\end{eqnarray}

The contribution referring to the regular part Eq. (\ref{37}) does not have an analytical expression and, therefore, requires a numerical evaluation. In the same way, we do for the deflection of light when subjected to a gravity field in the spacetime of an ABB which is given by $\delta\phi= \Delta\phi -\pi$.

\begin{figure}[h]
   	\centering
   	\includegraphics[width=\columnwidth]{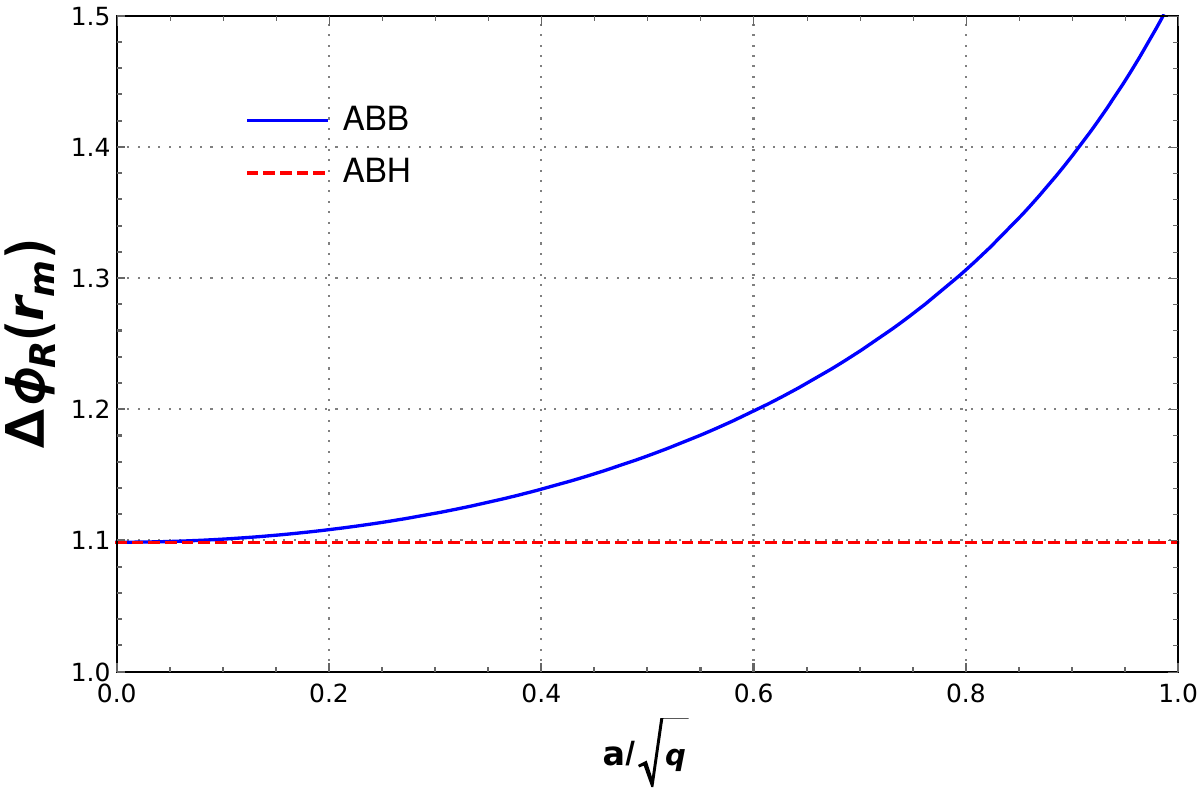}
   	\caption{Regular part of the integration of angular deviation.} 
   	\label{REGU}
   \end{figure}

\begin{figure}[ht]
   	\centering
   	\includegraphics[width=\columnwidth]{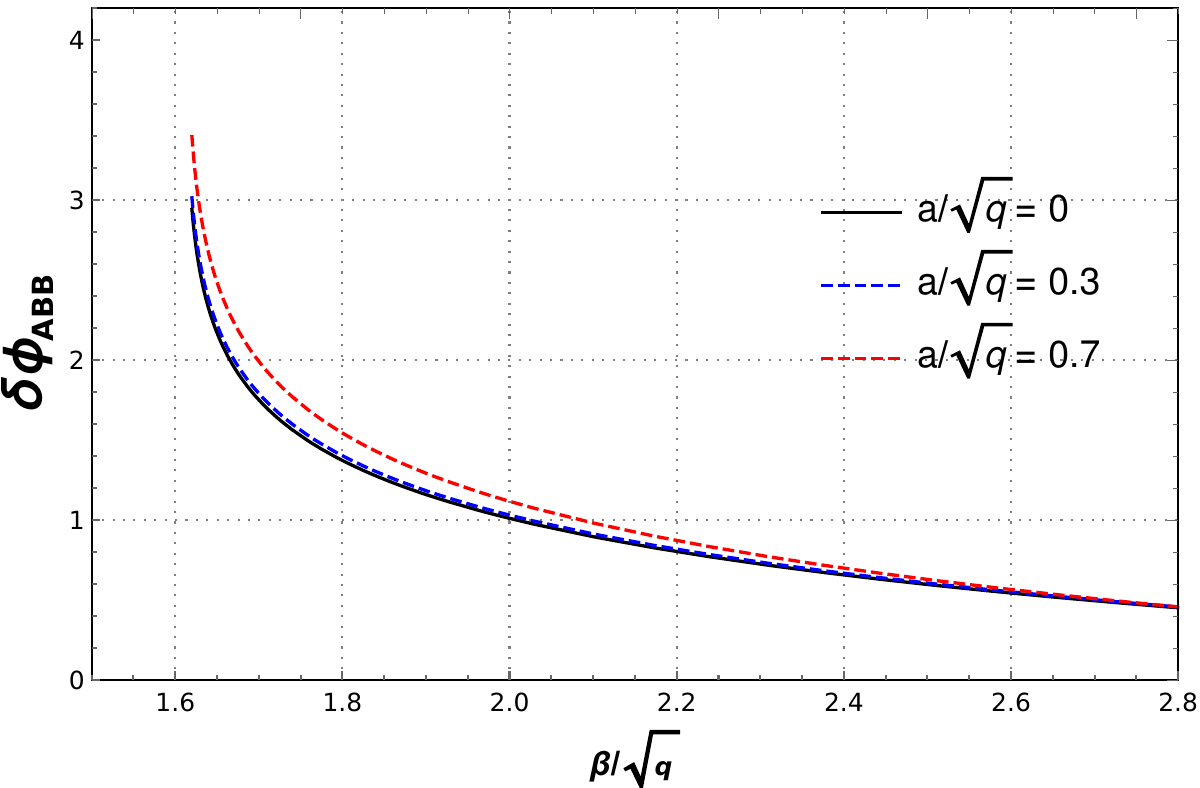}
   	\caption{Deviation of light as a function of the impact parameter for some values of throat radius $a$.} 
   	\label{DESVIO1}
   \end{figure}
 \begin{figure}[ht]
   	\centering
   	\includegraphics[width=\columnwidth]{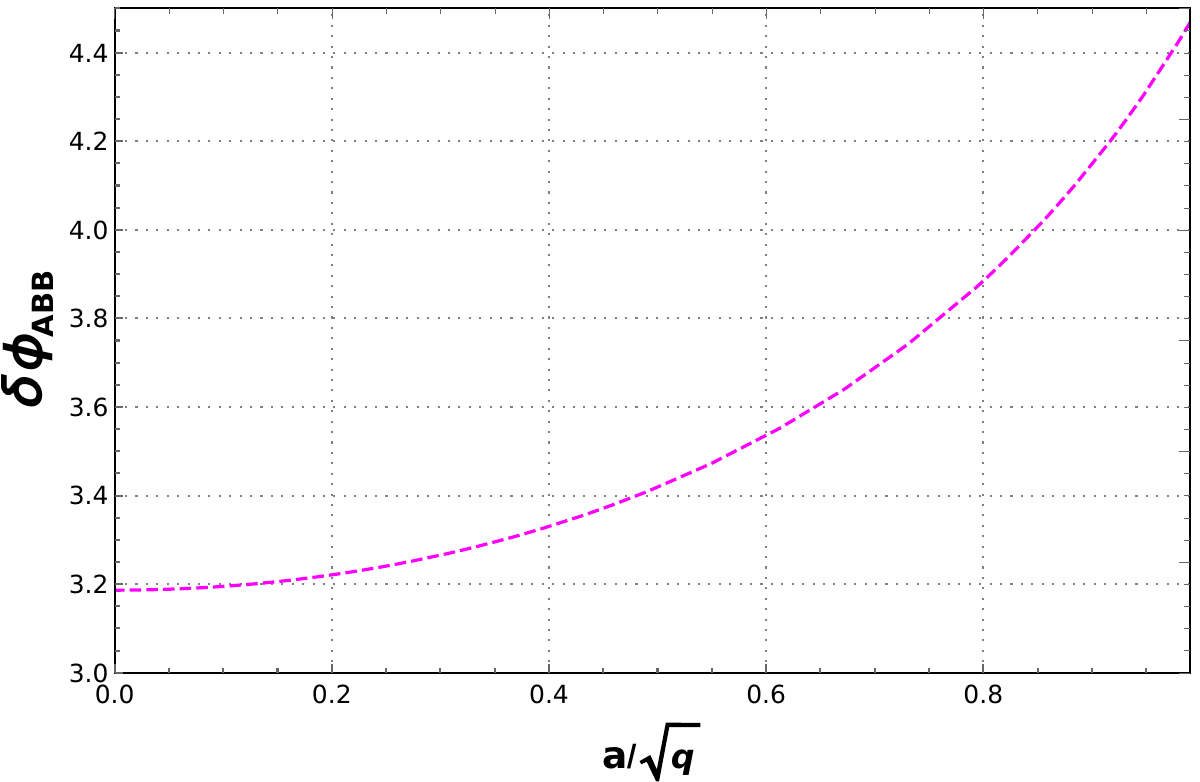}
   	\caption{Deviation of light as a function of $a/\sqrt{q}$ for $\beta/\sqrt{q}=\frac{3^{3/4}}{\sqrt{2}}+0.005$.}
   	\label{DESVIO2}
   \end{figure}

The panel in Fig. (\ref{REGU}) represents the behavior of the integration of the regular part of the light deviation, Eq. (\ref{37}), where the blue curve configures the spacetime of the ABB, and the red curve is the limit in which the ABB throat radius tends to zero. Then, spacetime starts to be represented by the ABH discussed in section \ref{sec3}.

The panel in Fig. (\ref{DESVIO1}) illustrates the total angular deviation $\delta\phi$ performed by light in the ABB spacetime as a function of the ratio $\beta/\sqrt{q}$, taking into account some values related to the radius of the ABB throat, with the black curve representing the deviation related to the ABH. Finally, in panel Fig. (\ref{DESVIO2}) we have the representation of the angular deviation of light in the ABB spacetime taking into account the ratio $\frac{\beta}{\sqrt{q}}=\frac{3^{3/4}}{\sqrt{2}}+0.005$.

\section{LENS EQUATION AND OBSERVABLES}\label{sec5}

This section studies the connection between the light deflection in the weak field regime, Eq. (\ref{26}), and the strong field regime, Eqs. (\ref{36}) and (\ref{37}), with $\delta\phi= \Delta\phi_R + \Delta\phi_D -\pi$, with the master equations of gravitational lensing in the spacetime of an ABB. Thus, we can construct physical quantities that can be theoretically observed. We consider lenses in the strong field regime as a starting point.

\begin{figure}[ht]
   	\centering
   	\includegraphics[width=\columnwidth]{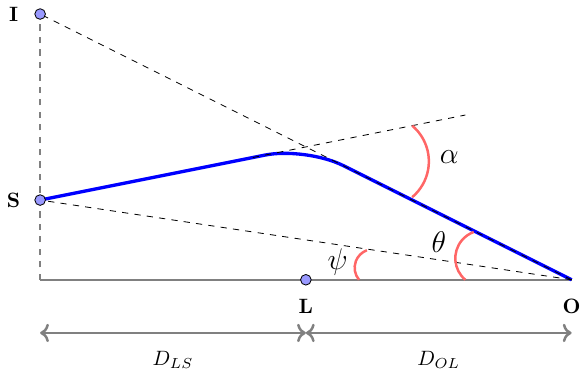}
   	\caption{Light angular deflection diagram.} 
   	\label{LENTE1}
   \end{figure}
   
In Fig. (\ref{LENTE1}) we have a panel that visually represents the lens diagram, where the light beam that is emitted through source \textbf{S} is deflected toward observer \textbf{O} due to the presence of the ABB located at \textbf{L}. The angular deflection of light is given by $\alpha$. The angular positions of the source and the image in relation
to the optical axis, $\overline{LO}$, are given, respectively, by $\psi$ and $\theta$. Thus, assuming that the source \textbf{S} is perfectly aligned with the lens \textbf{L}, this is the position in which the relativistic images should have the highest concentration \cite{8,9}. We then have the lens equation that relates the angular positions $\psi$ and $\theta$ defined as
\begin{eqnarray}\label{LT1}
\psi = \theta -\frac{D_\text{LS}}{D_\text{OS}}\Delta\alpha_n,
\end{eqnarray} 
where $\Delta\alpha_n$ is the deflection angle subtracted from all the loops made by the photons before reaching the observer, that is, $\Delta\alpha_n = \alpha -2n\pi$. In this approach, we use the following approximation for the impact parameter $\beta\approx\theta{D_\text{OL}}$. Thus, we can rewrite the expression for angular deflection using Eqs. (\ref{36}) and (\ref{37}) as
\begin{eqnarray}\label{LT2}
    \alpha\left(\theta\right)= -\frac{\Bar{a}}{2}\log\left(\frac{\theta{D_\text{OL}}}{\beta_c}-1\right) +\Bar{b},
\end{eqnarray} 
in which
\begin{eqnarray}\label{LT3}
    \bar{b}&=& -\bar{a}\log\left[\frac{\sqrt{3}|q|}{2(3|q|-\sqrt{3}a^2)}\right] + \Delta\phi_R -\pi, \\ \label{LT31}
    \bar{a}&=& \sqrt{\frac{\sqrt{3}|q|}{(\sqrt{3}|q|-a^2)}},
\end{eqnarray} being that $\beta_c=3^{3/4}\sqrt{|q|/2}$, and $\Delta\phi_R$ is defined in Eq. (\ref{37}).  

Regarding $\Delta\alpha_n$, we can perform the expansion of the angular deviation $\alpha(\theta)$ in relation to the position $\theta$ using the following boundary conditions $\theta=\theta^0_n$ and $\alpha(\theta^0_n)=2\pi{n}$:
\begin{eqnarray}\label{LT4}
\Delta\alpha_n= \frac{\partial\alpha}{\partial\theta} \Big|_{\theta=\theta^{0}_n}\left(\theta -\theta^{0}_n\right).
\end{eqnarray}

Applying such boundary conditions for the angular position $\theta$ in the angular deflection equation (\ref{LT2}), we have
\begin{eqnarray}\label{LT5}
    \theta^{0}_n= \frac{(1+e_n)\beta_c}{D_\text{OL}},  \qquad \text{with} \qquad e_n= e^{\frac{2(\bar{b}-2\pi{n})}{\bar{a}}}.
\end{eqnarray}

Therefore, substituting Eqs. (\ref{LT2}) and (\ref{LT5}) into Eq. (\ref{LT4}), we have
\begin{eqnarray}\label{LT6}
    \Delta\alpha_n= -\left[\frac{\bar{a}D_\text{OL}}{2\beta_c{e_n}}\right](\theta-\theta^0_n).
\end{eqnarray}

We can also construct a new relation between the angular positions by substituting Eq. (\ref{LT6}) into Eq. (\ref{LT1}). So, we have 
\begin{equation}\label{LT7}
    \theta_n \approx \theta^{0}_n + \left(\frac{2\beta_c{e_n}}{\bar{a}}\right) \frac{(\psi -\theta^0_n)D_\text{OS}}{D_\text{LS}D_\text{OL}}.
\end{equation}

We know that during the process of deflection of light the brightness of its surface is preserved. However, the position of the solid angle of the source is modified due to the presence of gravitational lensing. Thus, the total flux received by a lensed image is proportional to its magnification $\mu_n$, which is defined by $\mu_n=\Big| \frac{\psi}{\theta}\frac{\partial\psi}{\partial\theta}\mid_{\theta=\theta^{0}_n}\Big|^{-1}$. Therefore, using the expressions Eqs. (\ref{LT2}) and (\ref{LT4}) in the amplification equation, we have that
\begin{equation}\label{LT8}
    \mu_n= \frac{2e_n(1+e_n)}{\psi\bar{a}}\left(\frac{\beta_c}{D_\text{OL}}\right)^2\frac{D_\text{OS}}{D_\text{LS}}.
\end{equation}

Through the amplification expression above, we can see that an exponential decay occurs according to $n$, indicating that the brightness of the first image $\theta_1$ plays a dominant role over the others.  On the other hand, the presence of the factor $\left(\frac{\beta_c}{D_\text{OL}}\right)^2$
indicates that the amplification is always small. In the limit of $\psi\to 0$, the scenario in which the maximum alignment between the source, the lens, and the observer occurs, the amplification expression must diverge, thus maximizing the possibility of detecting relativistic images.

\subsection{Observables in the strong field limit}\label{sec51}

In the previous sections, we expressed the relativistic positions of the images, as well as their fluxes as a function of the expansion parameters $\bar{a},\bar{b}$, and $\beta_c$. Thus, we now do the reverse process, reconstructing the expansion coefficients and taking the observations as a starting point. Thus, we can investigate the characteristics of the object that is generating the gravitational lens and then compare this observation with theoretical predictions. The impact parameter may be written in terms of $\theta_\infty$ \cite{5},
\begin{equation}\label{LT9}
    \beta_c=D_\text{OL}\theta_\infty.
\end{equation}

From Bozaa's point of view \cite{5}, we assume that only the outermost image $\theta_1$ is resolved as a single image while the others are encapsulated in the boundary $\theta_\infty$. Thus, Bozza defined the following observables:
\begin{eqnarray}\label{LT10}
    s&=&\theta_1-\theta_\infty=\theta_\infty{e}^{\frac{2(\bar{b}-2\pi)}{\bar{a}}}, \\ \label{LT11}
    \tilde{r} &=& \frac{\mu_1}{\sum^\infty_{n=2}\mu_n}=e^{\frac{4\pi}{\bar{a}}}.
\end{eqnarray}

In the expressions referring to the observables above, Eqs. (\ref{LT10}) and (\ref{LT11}), $\textbf{s}$ represents the angular separation and $\tilde{r}$ the connection between the flow referring to the first image with the others. These forms can be inverted to obtain the expansion coefficients, and with this, we can compare observational data with the models studied.

\begin{figure}[ht]
   	\centering
   	\includegraphics[width=\columnwidth]{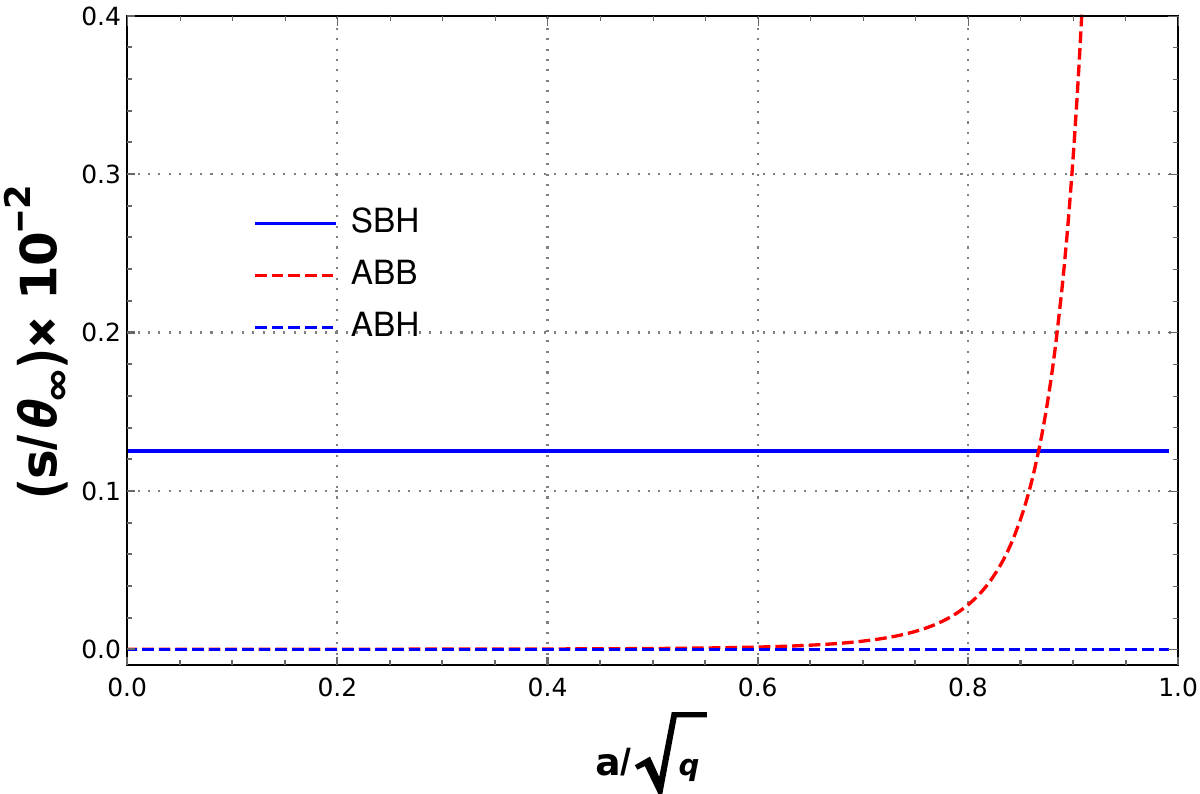}
   	\caption{Behavior of the ratio $s/\theta_\infty$.} 
   	\label{Sm}
   \end{figure}

In Fig. (\ref{Sm}), we have the representation of the ratio $s/\theta_\infty$ as a function of the relationship between the radius of the throat bounce and the square root of the magnetic charge $a/\sqrt{q}$. In Fig. (\ref{Sm}), we compare the behavior of this aforementioned ratio for the SBH (solid blue curve), ABH (dotted blue curve), and ABB (dotted red curve). 

   To simplify, we also constructed table \ref{TAB1} with some specific values. It is worth noting that, for the Schwarzschild solution, $s/\theta_{\infty}\sim 10^{-3}$. As we can observe, in the case of the ABH, $s/\theta_{\infty}\sim 10^{-7}$; this suggests that measuring the angular separation of an astrophysical analog of the ABH might be more difficult than in the Schwarzschild case. On the other hand, as we can observe, in the case of the ABB, in the limit $a/\sqrt{q}\to 1$, $s/\theta_\infty$ tends to larger values than the Schwarzschild case, which implies an improvement in the possibility of detection.

\begin{table}[ht]

\centering
\caption{Observables}

\label{TAB1}
\def\arraystretch{1.9}

\begin{tabular}{@{}lrrrr@{}}
\toprule

\textbf{$a/\sqrt{q}$} & \textbf{$s/\theta_\infty$}  \\

\midrule
0.10          & $7.59\times{10^{-7}}$        \\
0.30        & $1.42\times{10^{-6}}$        \\
0.50        & $5.37\times{10^{-6}}$         \\
0.70        & $5.20\times{10^{-5}}$        \\
0.80        & $2.80\times{10^{-4}}$        \\
0.90        & $3.10\times{10^{-3}}$         \\
0.99        & $2.42\times{10^{-1}}$         \\
\bottomrule

\end{tabular}
\end{table}

\begin{figure}[ht]
   	\centering
   	\includegraphics[width=\columnwidth]{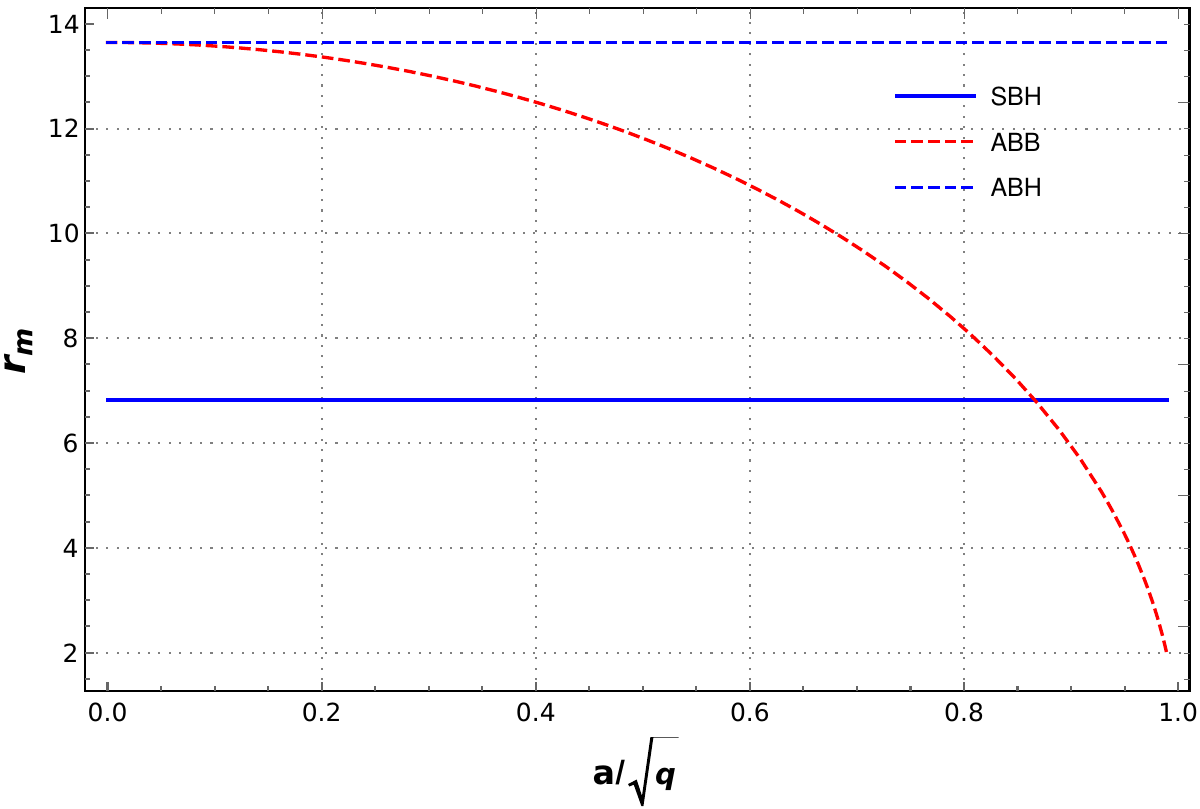}
   	\caption{Plot of $r_{m}=2.5\log_{10}\tilde{r}$ as a function of $a/\sqrt{q}$.} 
   	\label{MAG}
   \end{figure}

In Fig. (\ref{MAG}), we have the graphical representation for $2.5\log_{10}\tilde{r}$ also as a function of the ratio $a/\sqrt{q}$ where the colors of the curves follow the same scenarios mentioned previously. Thus, we can see that the the magnification decreases for increasing values of $a/\sqrt{q}$, which implies that the flux of the first relativistic image decreases relative to the others. In fact, it becomes smaller than in the Schwarzschild case when $a/\sqrt{q} \to 1$, meaning that the relativistic images are sharper than in the ABH and SBH cases.

\subsection{Observables in the weak field limit}\label{sec52}

The next step is to analyze the expressions referring to observables in the weak field regime, which means that the impact parameter is very large compared to the magnetic charge $\beta\gg{q}$, so the light beam does not form loops. Thus, we look at the expression for the angular deviation for the ABB in the weak field regime, Eq. (\ref{26}), and consider only terms up to second order in the impact parameter, we have that
\begin{equation}\label{LT12}
    \delta\phi \simeq \frac{\pi{a^2}}{4\beta^2}.
\end{equation}

For the expression referring to the angular position of the matter and the image Eq. (\ref{LT1}) considering that the system formed by (matter, compact object, and the observer) is in perfect alignment, we have $\psi=0$. Thus,
\begin{equation}\label{LT13}
    \theta= \frac{D_\text{LS}}{D_\text{OS}}\Delta\alpha_n,
\end{equation} where $\Delta\alpha_n$ is given by Eq. (\ref{LT12}). Therefore, substituting Eq. (\ref{LT12}) into Eq. (\ref{LT13}), we have the expression referring to the angular position of the Einstein ring,
\begin{equation}\label{LT14}
    \theta=\theta_E= \left[\left(\frac{\pi{a^2}}{4}\right)\left(\frac{D_\text{LS}}{D_\text{OS}D^2_\text{OL}}\right)\right]^{1/3}.
\end{equation}

Analyzing the above expression, Eq. (\ref{LT14}), it is clear that the position of the Einstein ring is being modified only by the presence of the ABB throat radius, and therefore, the magnetic charge itself does not play any role in this quantity, which makes this result have to coincide with that obtained for the Ellis-Bronnikov model \cite{r4a}.

Therefore, substituting Eq. (\ref{LT14}) into the expression below, we obtain the expression for the Einstein ring,
\begin{equation}\label{LT15}
    R_E= D_\text{OL}\theta_E= \left[\left(\frac{\pi{a^2}}{4}\right)\left(\frac{D_\text{LS}D_\text{OL}}{D_\text{OS}}\right)\right]^{1/3}.
\end{equation}

With the above expressions, Eqs. (\ref{LT14}) and (\ref{LT15}), we can calculate and possibly measure the observables $R_E$ and $\theta_E$ that depend on the bounce throat radius parameter $a$.

We consider for the procedure of our analysis in this weak field regime the lensing for a bulge star \cite{INTROB64}. To this end, we consider the following values for the parameters $D_\text{OL}=4Kpc$ and $D_\text{OS}=8Kpc$. The results computed for the observables using the data mentioned above do not need to be recalculated in the present work, as they should be the same as those presented in Tables 1 and 2 of Ref. \cite{INTROB64}.

\section{CONCLUSION}\label{sec6}
In the present work, we started by investigating the deflection of light when subjected to a gravitational field generated by the presence the gravitational analog of a magnetically charged ABH described in EsGB theory Ref. \cite{3,1} for the weak and strong field regimes. Applying the same analysis procedure to the model now describes the gravitational analog of a magnetically charged ABB. This ABB solution is supported by the combination of a phantom scalar field and a class of nonlinear electrodynamics in GR \cite{INTRO24}.

Particularly in relation to the weak field regime, we construct analytical expressions for the angular deviation in both spacetime Eqs. (\ref{10}) and (\ref{26}). For the ABB model, we can verify the consistency of the methodology by evaluating the limit at which the magnetic charge ceases to act $q\to{0}$; then, we can recover the same angular deviation related to the Ellis-Bronnikov spacetime \cite{r4a,r4b}. Regarding the strong field regime for the ABH, we construct analytical expressions for the integrals related to the divergent part Eq. (\ref{20}) and regular Eq. (\ref{21}), and then, we can obtain a closed expression for the total angular deviation. On the other hand, for the ABB model, we obtain an analytical expression for the divergent part of the integration, Eq. (\ref{36}), and we had to construct a graphical representation of the regular part,  Fig. \ref{REGU}. We also created a graphical representation to analyze the behavior of the angular deviation as a function of the ratio of the impact parameter very close to the photon orbit and the magnetic charge, Fig. \ref{DESVIO1}, as well as a representation of the angular deviation as a function of the ratio between the throat radius and the magnetic charge, Fig. \ref{DESVIO2}.

Finally, we analytically construct the gravitational lensing expressions for the ABB, and then from them, we write the equations for the observables in the strong field regime, Eqs. (\ref{LT10}) and (\ref{LT11}). Next, we write the expression for the Einstein ring position, Eq. (\ref{LT14}), and the other observable, Eq. (\ref{LT15}), for the weak field regime.


\section{ACKNOWLEDGMENTS}
M. S. would like to thank Funda\c{c}\~ao Cearense de Apoio ao Desenvolvimento Cient\'ifico e Tecnol\'ogico (FUNCAP) for the financial support. H. Belich would like to thank CNPq for financial support.


\end{document}